# Dynamic Acidity in Defective UiO-66

Sanliang Ling[a] and Ben Slater[a]

[a] Department of Chemistry, University College London, 20 Gordon Street, London, WC1H 0AJ, United Kingdom

E-mails: s.ling@ucl.ac.uk; b.slater@ucl.ac.uk

**Abstract**

The metal organic framework (MOF) material UiO-66 has emerged as one of the most promising MOF materials due to its thermal and chemical stability and its potential for catalytic applications. Typically, as-synthesised UiO-66 has a relatively high concentration of missing linker defects. The presence of these defects has been correlated with catalytic activity but characterisation of defect structure has proved elusive. We refine a recent experimental determination of defect structure using static and dynamic first principles approaches, which reveals a dynamic and labile acid centre that could be tailored for functional applications in catalysis.

## 1. Introduction

Intrinsic and extrinsic defects in metal organic frameworks[1-4] have emerged as an area of potential high importance for the translation of these materials to commercial applications.[5] For example, recent work has shown that different crystal faces of a MOF have profoundly different catalytic efficiency for biodiesel production[6] and mixed-ligand MOFs can have superior chemical and thermal stability to their end-members.[7] A major unsolved challenge is the characterization of defect structures and resolving their spatial distribution.[8-9] Thus far, there are a very small number of experimental papers that focus on defect structure characterisation in MOFs but there is growing canon of data.[10-11] Recently, Trickett et al. published a study that used X-ray diffraction (XRD) to shed light on the nature of missing linker defects in the MOF UiO-66.[12] Here, we report on new aspects of linker defect structure and behaviour using computational approaches that reveal a dynamic complexity that is invisible to time and spatially averaged XRD methods. We find evidence of shuttling protons within defective UiO-66 that may be important in understanding this material's catalytic efficacy.[13]

Perfect UiO-66 consists of a large $Zr_6$ metalloxalate cluster that is coordinated to 12 nearest neighbour $Zr_6$ metalloxalate clusters via 1,4-benzene-dicarboxylate ($BDC^{2-}$) linkers. Additionally, there are four 3-fold bridging OH groups (hereafter referred to as $\mu_3$-OH). The chemical nature of missing linker defects in UiO-66 has been under intensive debate in previous studies,[12, 14-19] and two questions remain unresolved: first, what is the chemical identity of the species that maintains the charge neutrality after the removal of the negatively charged $BDC^{2-}$ linker from the UiO-66 structure, and second, what is the defect structure? Distinct charge balancing chemical species have been suggested that could terminate the missing linker vacancies, including formate,[18] chloride[14, 16-17] and hydroxide.[12, 14-15] Formate has been excluded by $^1$H NMR experiments,[12, 16] and chloride has been discounted by energy-dispersive X-ray spectroscopy and coupled thermogravimetric and mass spectrometric analysis.[12] Therefore, in the current work, we focus on the structure of hydroxide terminated and charge compensated missing linker vacancies in UiO-66, which are expected to be particularly relevant to catalytic applications. Hydroxide species bound to metal sites are potential acid sites; for example, Zn-OH groups are believed to be responsible for the enhanced catalytic activity in defective MOF-5.[20]

In the recent X-ray diffraction work of Trickett et al., it was concluded the charge balancing hydroxide anions are stabilised by a hydrogen bond with a neighbouring $\mu_3$-OH group of UiO-66, while the two Zr atoms at the missing linker defect site are terminated by water molecules.[12] Through *ab initio* molecular dynamics (AIMD) simulations, we show this structural arrangement is thermodynamically disfavoured and the hydroxide anion resides on a Zr site. We further show that depending on the temperature and concentration (partial pressure) of extra-framework water molecules, dynamic acidity arises in defective UiO-66 due to a double proton transfer process involving two water molecules and one hydroxide anion, or a single proton transfer process involving a water molecule and a hydroxide anion. The dynamic acidity associated with missing linker vacancies in UiO-66 could be used to engineer catalytic active centres in this and similar materials.

All periodic density functional theory (DFT) calculations, including geometry/cell optimisations and AIMD simulations, have been performed using the CP2K code.[21-22] We have used both gradient corrected and hybrid density functional methods including dispersion interactions, informed from previous work on MIL-53 type MOF materials,[23-24] which gave very good agreement with experimental structural data and calorimetric data. More details of the calculations are included in the Supporting Information (SI) including a sample input.

## 2. Results

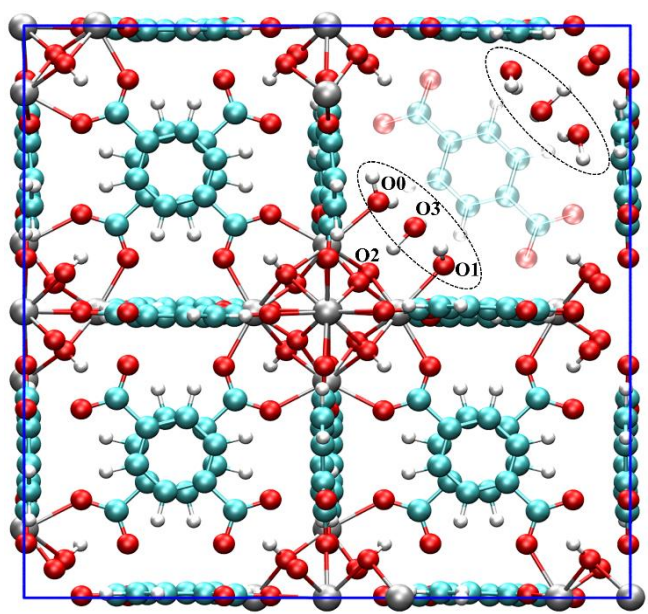

**Figure 1:** The unit cell of hydroxylated UiO-66 featuring one missing BDC$^{2-}$ linker. The defect structure shown (see dotted ellipsoid) was optimised from the experimental defect structure proposed by Trickett et al.[12] The BDC$^{2-}$ linker that is co-planar to the missing BDC$^{2-}$ linker is set to semi-transparent to emphasise the defect. Several oxygen atoms are labelled to aid the discussion. Colour code: C, cyan; O, red; H, white; Zr, grey.

Figure 1 shows a model of the defective structure of UiO-66, in which there is one missing BDC$^{2-}$ linker defect per unit cell (4.2% linker vacancy concentration). A 10% defect incidence has been widely reported experimentally (equivalent to two missing linkers per cell),[12, 25] but since individual defect centres are well separated from each other, the main findings also apply to UiO-66 samples with higher missing linker defect concentrations. In UiO-66, each $BDC^{2-}$ linker lies along a face diagonal bridging two $Zr_6$ clusters and hence each $BDC^{2-}$ vacancy creates two defect centres with four notionally under-coordinated Zr sites. In the recent work by Trickett et al.,[12] it was proposed that the missing $BDC^{2-}$ linker defect is charge balanced by two hydroxide anions that are hydrogen bonded to two $\mu_3$-OH groups in the parent UiO-66 material, while the four under-coordinated Zr atoms are bound to atmospheric physisorbed water, as depicted in Figure 1. The suggested binding of a bare hydroxide anion with the $\mu_3$-OH of the $Zr_6$ cluster is unusual and so we sought to examine this motif using periodic density functional theory methods.

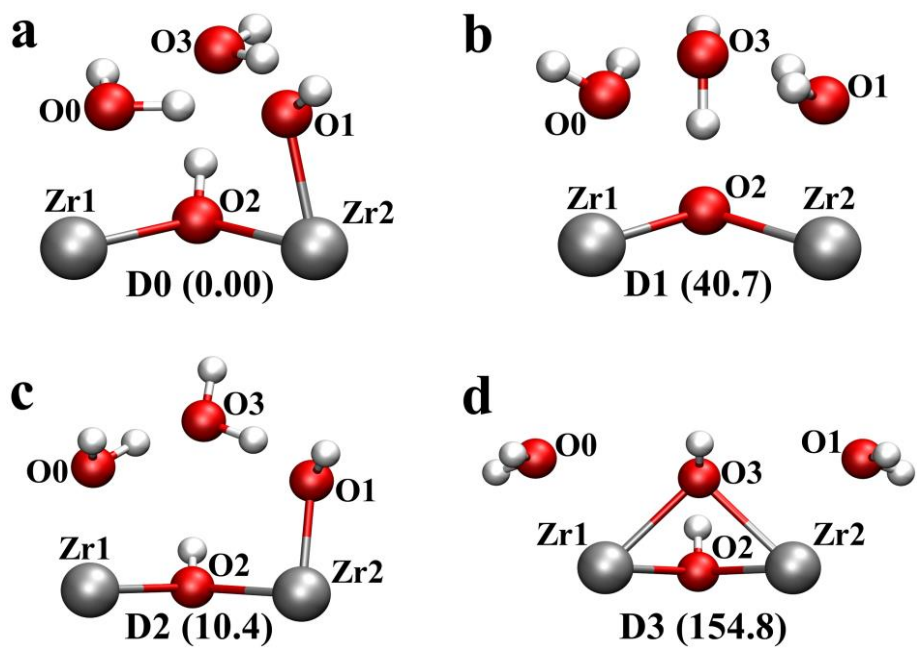

**Figure 2**: Four possible local geometries for missing linker defect structures in UiO-66. Two pairs of under-coordinated Zr centres are created with a single linker vacancy that are compensated by two hydroxide species and four water molecules, hence one hydroxide and two water molecules per Zr pair, as depicted. The relative energies are given in brackets with respect to defect configuration D0 (obtained at the PBE0+D3 level of theory in kJ/mol per defect centre). The colour code is identical to that used in Figure 1.

Starting from a configuration resembling the geometry proposed by Trickett et al.,[12] we found that upon structural relaxation, the proton belonging to the $\mu_3$-OH spontaneously transferred to the charge balancing hydroxide anion to form a water molecule. The resulting configuration is pictured as D1, see Figure 2b (note that the montage shows one of the two defect centres present in the configuration but the two defects are approximately symmetric). The observation is explicable by consideration of the strong Coulomb repulsion between the bare and $\mu_3$-OH anions, which is thermodynamically unfavourable.

**Table 1**: Calculated bond lengths and interatomic distances (in Å) of four defect structures in comparison with experiment. All data were obtained at PBE0+D3 level of theory.

| Distances | D0 | D1 | D2 | D3 | Expt[a] |
|---|---|---|---|---|---|
| Zr1⋯O0 | 2.30 | 2.35 | 2.27 | 2.40 | 2.20/2.28[b] |
| Zr2⋯O1 | 2.14 | 2.32 | 2.08 | 2.40 | 2.20/2.28[b] |
| O2⋯O3 | 2.76 | 2.49 | 2.94 | 2.26 | 2.787 |

[a] Taken from Trickett et al. at 200 K.[12]
[b] Two sets of Zr atoms at similar positions (separated by 0.165 Å) were resolved in the experiment, and hence the two reported distances are listed for comparison.

Next, we constructed several additional defect configurations, which differ from D1 in the initial position of the charge balancing hydroxide anion and in the local hydrogen bonding network involving the three oxygen atoms at O0, O1 and O3 positions, see Figures 1 and 2. To explore the potential energy surface more comprehensively, we performed AIMD calculations. A total of 63 configurations were extracted from three AIMD trajectories at 300 K (see Supporting Information for more details) and optimised at an effective temperature of 0 K. From the 69 optimised configurations (including 6 manually constructed configurations), we show four representative structures (including defect configuration D1) along with their relative stabilities with respect to the most stable defect geometry identified, D0, in Figure 2, and we show pertinent bond lengths and interatomic distances of the four defect structures determined at the hybrid PBE0+D3 (incorporating van der Waals interactions that also take into account three-body, dispersive triple-dipole terms[26]) level of theory in Table 1.

The most stable defect configuration D0 exhibits a neutral water molecule coordinated to Zr1 and a hydroxide anion bonded to Zr2, with the oxygen atom at O3 position belongs to a neutral water molecule. This structure is stabilised by six hydrogen bonds (H-bonds). The hydrogen atoms covalently bonded to O0 and O2 form three H-bonds with O3, and O1 forms two H-bonds with hydrogen atoms from O0 and O3. In the undefective cell, the Zr-O(BDC$^{2-}$) bond length is predicted to be 2.22 Å at PBE0+D3 level of theory, which is in good agreement with the shorter Zr-O distance, i.e. 2.20 Å, reported by Trickett et al. at 200 K.[12] In defect configuration D0, the calculated O2⋯O3 distance of 2.76 Å is in excellent agreement with the recently reported distance of 2.730(6) Å by experiment.[12] We also find two distinct Zr⋯O distances of 2.14 Å and 2.30 Å in static calculations, corresponding to Zr-OH and Zr⋯H$_2$O respectively, which are in reasonable agreement with reported experimental distances of 2.20 Å and 2.28 Å obtained at 200 K (uncertainties in the experimental distances were not reported for the 200 K data but they are expected to be significant). The two Zr⋯O distances obtained from theory differ by 0.16 Å because the anionic hydroxide species binds more strongly with the Zr cation than the neutral water molecule.

Configuration D1 is similar to the arrangement proposed by Trickett et al.,[12] but in D1, the hydrogen atom bonded to O2 spontaneously transfers to O3 and forms a neutral water molecule which is hydrogen bonded to three neighbouring oxygen atoms, including O0, O1 and O2 ($\mu_3$-O). This is because there is no minimum on the potential energy surface for the geometry proposed by Trickett et al.[12] However, configuration D1 is higher in energy than D0 by 40.7 kJ/mol per defect centre, and therefore, it is very unlikely that this defect configuration could occur at relevant temperatures. We note that our optimised distance between the two oxygen atoms in defect configuration D1, i.e. $\mu_3$-O (O2) and O3, is 2.5 Å which is notably different from the experimental measurement of 2.8 Å by Trickett et al. Because defect configuration D1 is a high energy minimum and its structure is not compatible with that resolved using XRD, we can eliminate D1.

Defect configuration D2 is similar to D0 in structure and it is found to be slightly higher in energy than D0 by 10.4 kJ/mol per defect centre, indicating D2 is competitive with D0 at elevated temperatures. Again, the hydroxide is bonded to Zr rather than hydrogen bonded to the $\mu_3$-OH. D2 and D0 differ only in the local hydrogen bonding network. Moreover, among the 63 configurations optimised from the three MD trajectories, a total of 34 structures fall into a small energy window of 5.0 kJ/mol (~2 $k$T at 298 K) less stable than D0. These 34 structures differ from D0 in the hydrogen bonding networks formed by the two hydroxide anions and four water molecules adjacent to the four under-coordinated Zr sites. Crucially however, none of the structures exhibit a bare hydroxide hydrogen bonded to $\mu_3$-OH; all minima feature a single water bound to a Zr, a hydroxide species bound to a Zr and a bridging water molecule in the O3 position. The existence of so many conformations which lie close to those of the global minimum is not surprising as the hydrogen bonding potential energy surface is relatively shallow and complex; even the water dimer has 6 minima which have relative energies within 4.2 kJ/mol (< 2 $k$T at 298 K).[27]

We have also considered an additional defect configuration D3 in Figure 2d, which features a $\mu_2$-OH between the two Zr atoms as charge balancing anion. After geometry relaxation, no H-bond is evident in this configuration (within a 2.5 Å cutoff), but the $\mu_2$-OH is preserved, also reminiscent of the configuration proposed by Trickett et al.[12] However this configuration was found to be higher in energy than D0 by 154.8 kJ/mol per defect centre. Therefore, clearly this configuration will not be present in experimental samples at temperatures relevant for catalysis.

Our calculations suggest that multiple low-energy defect configurations will co-exist, and therefore, it is expected that kinetic factors, including temperature, will play a significant role in determining the definitive defect structures that are present in the defective lattice. To get a better understanding of the defect structure at a finite temperature, we performed AIMD simulations for 30 ps (60,000 MD post equilibration steps with a timestep of 0.5 fs) at 300 K, and we show how the relevant distances (e.g. Zr1···O0, Zr2···O1 and O2···O3) evolve as a function of time in Figure 3.

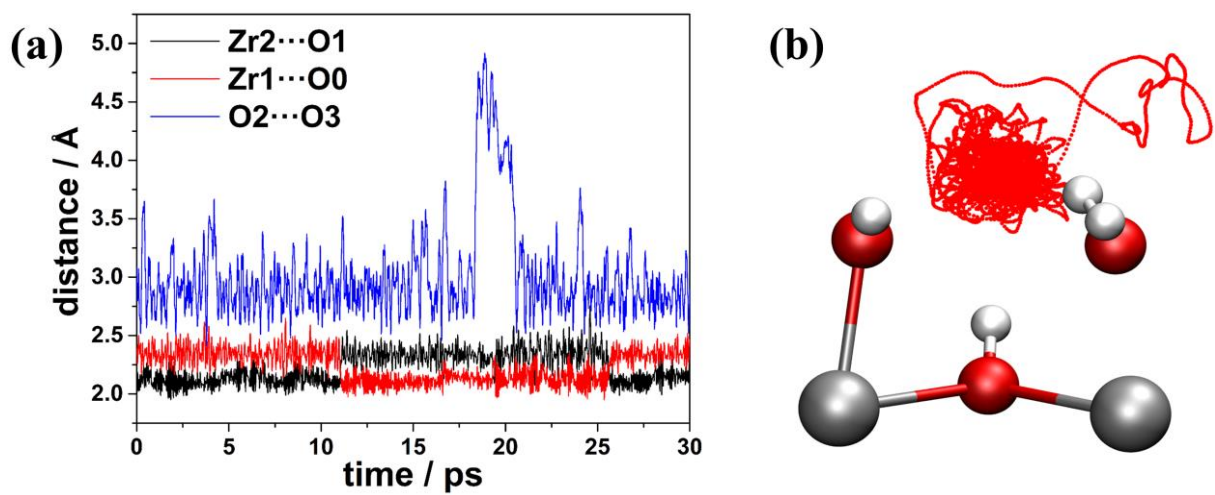

**Figure 3**: (a) Changes of distances (in Å) and (b) trajectory of the O3 atom (represented by red dotted line; colour code is the same as Figure 1) during the AIMD simulation at 300 K.

In Figure 3a, the O2···O3 distance approaches 5 Å at ~20 ps, indicating the neutral water molecule at O3 site is loosely bound (confirmed by inspection of Figure 3b which shows the trace of O3 over the 30 ps AIMD simulations), suggesting that this water molecule can leave the defect centre and become absorbed somewhere else in the bulk of the material or even enter into the atmosphere via the external surface of the sample. Indeed, we found that in defect configuration D0 (see Figure 2a), the binding energy of the water molecule at the O3 site is 86.9 kJ/mol per water molecule, which is considerably smaller than that of the water molecule at the Zr1 site (161.5 kJ/mol per water molecule). Another very interesting observation is that the two Zr···O distances (Zr1···O0 and Zr2···O1) change significantly and in complement to each other. The shorter Zr···O distance corresponds to Zr-OH, and the longer Zr···O distance is associated with Zr···H$_2$O. From analysis of the trajectory from the AIMD simulations, we found that the change in the two Zr···O distances is related to two simultaneous proton transfer processes facilitated by O3. Taking defect configuration D2 (see Figures 2c and 4a) as an example, a proton transfers from O3 to O1 and forms a neutral water molecule at O1 site, and the Zr2···O1 distance increases by ~0.2 Å (black solid line at ~11 ps in Figure 3a). At the same time, another proton transfers from O0 to O3, which leaves a hydroxide anion at O0 site, and the Zr1···O0 distance decreases by ~0.2 Å (red solid line at ~11 ps in Figure 3). After ~15 ps, similar double proton transfer happens again, and the two Zr···O distances change back to the previous state, indicating that the proton transfer processes are reversible. Based on accumulated statistics from multiple AIMD runs, the proton transfer happens on a relatively short time scale (10~15 ps), so an XRD experiment would see an average of the two Zr···O distances shown in Figure 3a. Averaging the two Zr···O distances over the trajectory of the whole AIMD simulation, we obtain a mean Zr···O distance of 2.23 Å, which is in excellent agreement with the experimental Zr···O distance of 2.24(3) Å at the same temperature of 300 K.[12] Note, we also performed AIMD simulations starting from different starting defect configurations (see Supporting Information) and these simulations gave qualitatively consistent results with those described here.

Additional AIMD simulations were performed at 100 K, 500 K and 700 K (see Supporting Information). At 100 K, we did not see proton transfer within the 10 ps window of simulation that we considered, which is expected since activated runs at 300 K show events on a 10~15 ps interval. At 500 K, we found the water molecule at O3 site diffused into the pore of the material after ~4 ps (and did not return in a run of length 10 ps). In addition, we found reversible proton transfer took place more often, at a frequency of one proton transfer per ps, ergo the proton transfer rate was enhanced after the water molecule departed from O3 site. These findings show that temperature plays a very important role in determining the dynamic behaviour of the missing linker defects in UiO-66. At low temperature, proton transfer and the motion of physisorbed water molecules at O3 site are suppressed whilst at high temperature, the water molecule at O3 site has enough kinetic energy to diffuse within the pore of UiO-66 and proton transfer is dramatically enhanced. At 700 K, we found the water molecule at O3 site diffused into the pore of the material during equilibration, and the water molecule at Zr site desorbed and diffused into the pore of the material after ~1 ps of production run (and did not return in a run of length 10 ps), leaving behind a hydroxide anion bonded to one of the under-coordinated Zr atoms and a bare Zr site.

In addition to hydroxide, Trickett et al. also considered propoxide as charge balancing anion, by synthesising UiO-66 from zirconium propoxide instead of zirconium oxychloride, and they concluded the propoxide anionic oxygen sits at the O3 site and hydrogen bonds to $\mu_3$-OH.[12] For comparison with the case of hydroxide, we also performed static and AIMD simulations at 300 K with propoxide as the charge balancing anion. We found the defect centre resembles that when hydroxide is the counterion; a proton transfers from water to the propoxide anion to form a neutral propanol molecule which has an oxygen atom at the O3 site, leaving behind a hydroxide anion bonded to a Zr atom, and a water molecule coordinated to the second under-coordinated Zr atom. More detailed discussions are included in the Supporting Information. These results appear to unambiguously show the charge compensating anions are bonded to the Zr metal site and not coordinated to the $\mu_3$-OH as previously proposed.[12] However, we emphasise the oxygen positions determined through first-principles calculations are compatible with those identified by XRD, only the position of the hydrogen atoms differ.[12]

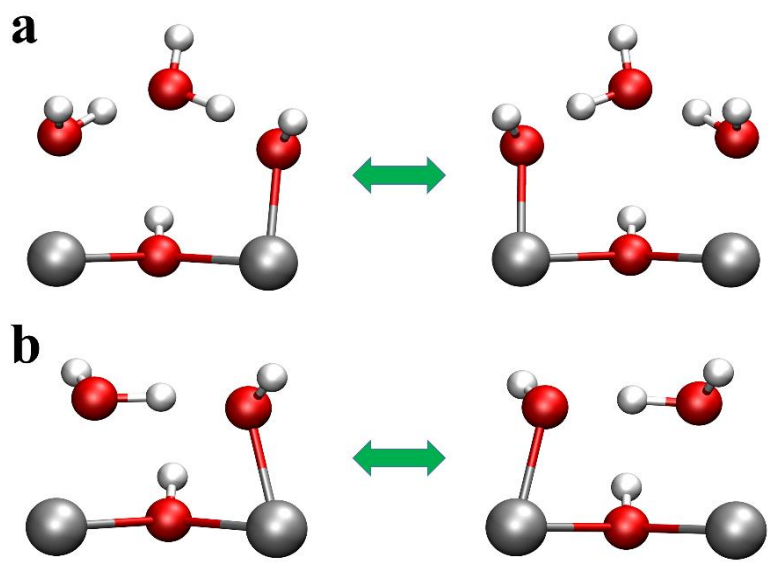

**Figure 4**: Proton transfer between O0 and O1 (a) involving O3, and (b) without O3.

Having established the detailed structure of the linker defect centre, we next examined the transition barriers of the proton transfer processes, as shown in Figure 4. Taking defect configuration D2 as the exemplar, we show the initial and final states of the double proton transfer processes between O0⋯O3 and O3⋯O1 in Figure 4a. To estimate the transition barrier of the double proton transfer processes, we took a linear interpolation of the Cartesian coordinates of the initial and final geometries to represent the reaction pathway (using a total of seven intermediate images), and performed geometry optimisation on the hydrogen atoms for the intermediate images on the reaction pathway. The transition barrier is estimated to be 27.5 kJ/mol per defect centre at the PBE0+D3 level of theory, indicating relatively facile proton transfer. As we showed earlier, the water molecule at O3 site can readily vacate its position and diffuse into the pore of the material, which is accompanied by relaxation of the hydroxide and water molecule at the defect centre, see Figure 4b. The result is that the O0⋯O1 distance decreases by ~1.0 Å; the O0⋯O1 distance in Figure 4b is 2.52 Å, compared to 3.57 Å in the case of defect configuration D2 (see Figures 2c and 4a). Evidently the proton can transfer from O0 to O1 aided by the reduced O0⋯O1 separation. We estimate the transition barrier of this single proton transfer process in the same manner as the O3 mediated case, which is found to be only 6.5 kJ/mol per defect centre (at the PBE0+D3 level of theory), 21.0 kJ/mol per defect centre lower than the two-step O3 mediated case. These estimated barriers are consistent with the AIMD simulations at 500 K, where we found that after the water molecule at O3 site diffused into the pore of the material and the proton transfer between O0 and O1 took place very frequently at a rate of one event per ps. In the presence of physisorbed water at O3, a rate of 1 event per ~15 ps for the double proton transfer is seen at 300 K. We note the configuration shown in Figure 4b resembles the proton topology of another $Zr_6$-based MOF material, NU-1000, as suggested by Cramer and co-workers.[28-29]

## 3. Discussions

Whilst the mobility of the water molecule at the O3 site and proton transfer processes in the vicinity of a missing linker defect in UiO-66 have a significant effect on the defect structures, the dynamic behaviour of the defects uncovered here will also affect the properties of the material. One of the applications that has been considered for metal-organic framework materials like UiO-66 is heterogeneous catalysis. In the case of undefective UiO-66, $\mu_3$-OH could be active as a Brønsted acid site, and it may be involved in applications like ammonia capture,[30] which can be improved by increasing the density of Brønsted acid sites.[31] In the case of UiO-66 with missing linker defects, intuitively, under-coordinated Zr atoms could be considered to be Lewis acids. However, from the AIMD simulations and extensive experimental studies on the nature of the missing linker defect in UiO-66, it is clear that the majority of the Zr atoms at defect sites will not be naked but terminated with water or hydroxide and therefore they cannot function as Lewis acids. On the other hand, the presence of charge balancing hydroxide anions due to missing $BDC^{2-}$ linkers is likely to increase the number of Brønsted acid sites within the material.

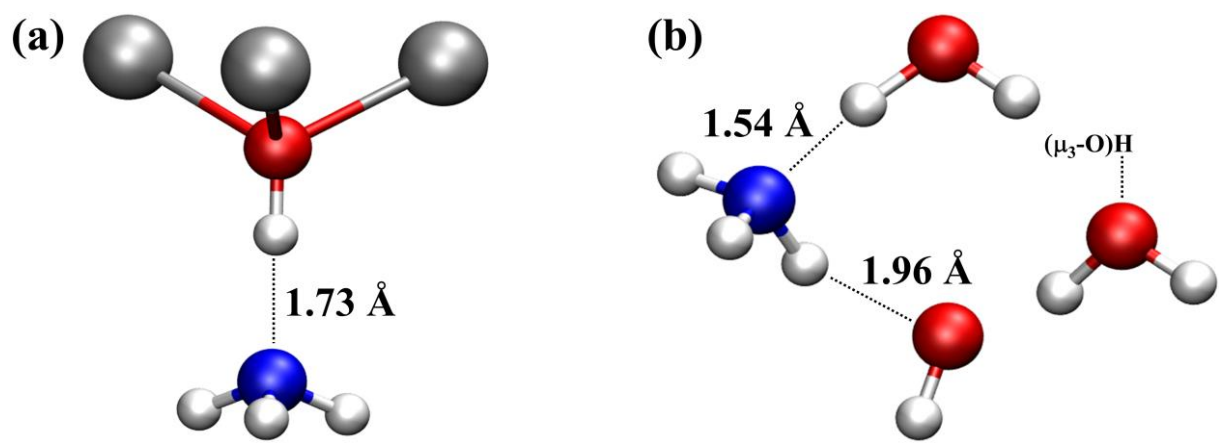

**Figure 5**: Local structures of ammonia adsorption in the (a) undefective $\mu_3$-OH centre and (b) defective regions of UiO-66. Relevant H-bond distances (in Å) are indicated to aid the discussion. The water molecule and hydroxide anion next to ammonia molecule in (b) are coordinated to two Zr atoms at a defect centre, respectively. The water molecule on the right hand of (b) is hydrogen bonded to a $\mu_3$-OH. Colour code: N, blue; others are identical to those used in Figure 1.

To estimate the Brønsted acidity of UiO-66, we calculated the binding energy of ammonia ($NH_3$) molecule, one of the standard molecules to quantify the strength of Brønsted acidity, in the defective and undefective regions of UiO-66. We find that the binding energy of

NH$_3$ in the undefective region (see Figure 5a) is 75.8 kJ/mol per NH$_3$ molecule but 110.1 kJ/mol in the defective region (see Figure 5b), clearly demonstrating enhanced binding at the defect centre. In the undefective part of the lattice, the NH$_3$ molecule forms a single O-H$\cdots$N hydrogen bond with the $\mu_3$-OH of UiO-66. However, in the defective region, because of the presence of extra-framework water molecules and the charge balancing hydroxide anion, the NH$_3$ molecule forms two hydrogen bonds; one O-H$\cdots$N hydrogen bond with Zr$\cdots$H$_2$O and another N-H$\cdots$O hydrogen bond with Zr-OH. The O-H$\cdots$N hydrogen bonding distance at the defect centre is substantially shorter than that found in the perfect region by 0.2 Å. The small O$\cdots$N separation of ~2.6 Å is expected to give a low transition barrier for proton transfer from water to ammonia at the defect centre and smaller than that from $\mu_3$-OH to ammonia in the perfect region, suggesting higher acidity and lability at the defect centre.

Another consequence of the larger binding energy of ammonia in the defective region is that the residence time of the ammonia molecule at the defect centre will be greater than that in the undefective region, and hence the probability that ammonia can receive a proton, is also enhanced. The stronger binding at the defect centre (in comparison to the perfect lattice) is due to the presence of polar water molecules and the hydroxide anion. Ammonia has a permanent electric dipole moment of 1.47 D,[32] and hence has a stronger electrostatic attraction to the polar defect centre. Confirmation of the enhanced binding can be found in a recent experimental study of ammonia uptake in UiO-66 with variously functionalised BDC$^{2-}$ organic linkers,[31] where it was found that in dry air conditions, ammonia uptake in UiO-66-OH is much higher than in UiO-66 and UiO-66-(OH)$_2$. By structural inspection, the latter two materials are less polar compared with UiO-66-OH. This property may be considered in future defect engineering of similar MOFs for gas separation[33] and ammonia capture[30-31] and clearly emphasises the potential for participating in catalytic processes. In more general terms, our findings also suggest that the polar defect centres could also help trap larger molecules and enhance the probability of proton transfer from the $\mu_3$-OH Brønsted acid site.

The AIMD simulations at 300K and 500 K show that the water molecule hydrogen bonded to $\mu_3$-OH at a defect centre can diffuse away. At higher temperatures, it can be anticipated that the second water molecule coordinated to the under-coordinated Zr atom may desorb and diffuse into the pore. Indeed, this is observed during the AIMD simulation at 700 K (see Figure S1c in the Supporting Information). This results in an open Zr site, a Lewis acid, adjacent to a hydroxide (bonded to the second under-coordinated Zr atom), a Lewis base. This is a textbook example of the so-called frustrated Lewis pair, which has been demonstrated to show high catalytic activity towards a range of chemical reactions, e.g. activation of $CO_2$.[34-35] The presence of frustrated Lewis pair sites in defective UiO-66 would be expected to increase the catalytic activity of the material, and the prospect of an experimental verification of this prediction is tantalising.

Finally, missing linker vacancies are one example of "point" defects within UiO-66, another is a missing $Zr_6$ metal cluster (where the organic linkers connected to the $Zr_6$ metal cluster are also missing),[8] which can be regarded as a cluster of missing linker vacancies plus a missing $Zr_6$ metal cluster, because the terminations of the under-coordinated Zr sites will be the same in both cases. These defects give rise to mesoscopic pores and that will resemble the external surface but in the crystal interior. Indeed, while this study examined vacancies in the bulk of the material, it is reasonable to suppose that at the external surfaces of an UiO-66 nanoparticle, there will be similar under-coordinated Zr sites terminated by charge balancing hydroxide anions and water molecules. Therefore, it is expected that the dynamic acidity and potential for frustrated Lewis pair sites found in the crystal interior could be present on the external surfaces of UiO-66 samples.

## 4. Conclusions

To conclude, extensive static and *ab initio* molecular dynamics simulations on UiO-66 with missing organic linkers have been performed, and the results demonstrate that charge balancing hydroxide anions are bonded to under-coordinated Zr sites, creating potential acid centres. Crucially, we further show that the defect structures exhibit strong dynamic behaviour associated with rapid proton transfer involving the hydroxide anion and extra-framework physisorbed atmospheric water molecules. The chemical species bonded to the Zr atoms at the defect centre show a fluxionality, alternating between hydroxide and water, a process that is mediated by proton transfer. The defect centres show increased acidity and enhanced trapping properties and a source of highly mobile protons. Under highly activated conditions, frustrated Lewis pair sites may form. All of the aforementioned properties arise because of the presence of defects in the UiO-66 material, which undoubtedly confers the potential for increased catalytic functionality and for tailoring the functional behaviour of this material. UiO-66 is an atypical MOF because it contains a high incidence of defects that allows the field to unambiguously chart the connection between defect presence and changes to properties and to characterise defect structure. It is tempting to suggest that similar defects may exist in a wide range of MOFs but that their concentration is simply much lower and therefore less amenable to detection by experimental approaches. Clearly further work is needed to characterise defects, to predict defect formation and incidence and to assess how defects influence properties, including reactivity.

**Acknowledgements**

We thank Greig Shearer for useful discussions on defects in UiO-66. We also thank Christopher Trickett for providing additional experimental data and discussion. This work is supported by EPSRC (EP/K039296/1). Through our membership of the UK's HPC Materials Chemistry Consortium, which is funded by EPSRC (EP/L000202), this work made use of the facilities of HECToR and ARCHER, the UK's national high-performance computing service, which is funded by the Office of Science and Technology through EPSRC's High End Computing Programme. Additionally, this work also made use of the resources awarded under the PRACE programme DECI-12 NANOMOF.

# References

1. Zacher, D.; Schmid, R.; Woll, C.; Fischer, R. A., *Angewandte Chemie International Edition* **2011**, *50*, 176-99.
2. Sholl, D. S.; Lively, R. P., *Journal of Physical Chemistry Letters* **2015**, 3437-3444.
3. Fang, Z.; Bueken, B.; De Vos, D. E.; Fischer, R. A., *Angewandte Chemie International Edition* **2015**, *54*, 7234-54.
4. Canivet, J.; Vandichel, M.; Farrusseng, D., *Dalton Transactions* **2016**, DOI: 10.1039/c5dt03522h.
5. Gascon, J.; Corma, A.; Kapteijn, F.; Llabrés i Xamena, F. X., *ACS Catalysis* **2014**, *4*, 361-378.
6. Liu, Y.; Liu, S.; He, D.; Li, N.; Ji, Y.; Zheng, Z.; Luo, F.; Liu, S.; Shi, Z.; Hu, C., *Journal of the American Chemical Society* **2015**, *137*, 12697-703.
7. Kalidindi, S. B.; Nayak, S.; Briggs, M. E.; Jansat, S.; Katsoulidis, A. P.; Miller, G. J.; Warren, J. E.; Antypov, D.; Corà, F.; Slater, B.; Prestly, M. R.; Martí-Gastaldo, C.; Rosseinsky, M. J., *Angewandte Chemie* **2015**, *127*, 223-228.
8. Cliffe, M. J.; Wan, W.; Zou, X.; Chater, P. A.; Kleppe, A. K.; Tucker, M. G.; Wilhelm, H.; Funnell, N. P.; Coudert, F.-X.; Goodwin, A. L., *Nature Communications* **2014**, *5*, 4176.
9. Ma, M.; Gross, A.; Zacher, D.; Pinto, A.; Noei, H.; Wang, Y.; Fischer, R. A.; Metzler-Nolte, N., *CrystEngComm* **2011**, *13*, 2828.
10. Chizallet, C.; Lazare, S.; Bazer-Bachi, D.; Bonnier, F.; Lecocq, V.; Soyer, E.; Quoineaud, A. A.; Bats, N., *Journal of the American Chemical Society* **2010**, *132*, 12365-77.
11. Mondloch, J. E.; Katz, M. J.; Isley, W. C., 3rd; Ghosh, P.; Liao, P.; Bury, W.; Wagner, G. W.; Hall, M. G.; DeCoste, J. B.; Peterson, G. W.; Snurr, R. Q.; Cramer, C. J.; Hupp, J. T.; Farha, O. K., *Nature Materials* **2015**, *14*, 512-6.
12. Trickett, C. A.; Gagnon, K. J.; Lee, S.; Gandara, F.; Burgi, H. B.; Yaghi, O. M., *Angewandte Chemie International Edition* **2015**, *54*, 11162-7.
13. Vermoortele, F.; Bueken, B.; Le Bars, G.; Van de Voorde, B.; Vandichel, M.; Houthoofd, K.; Vimont, A.; Daturi, M.; Waroquier, M.; Van Speybroeck, V.; Kirschhock, C.; De Vos, D. E., *Journal of the American Chemical Society* **2013**, *135*, 11465-8.
14. Katz, M. J.; Brown, Z. J.; Colon, Y. J.; Siu, P. W.; Scheidt, K. A.; Snurr, R. Q.; Hupp, J. T.; Farha, O. K., *Chemical Communications* **2013**, *49*, 9449-51.
15. Valenzano, L.; Civalleri, B.; Chavan, S.; Bordiga, S.; Nilsen, M. H.; Jakobsen, S.; Lillerud, K. P.; Lamberti, C., *Chemistry of Materials* **2011**, *23*, 1700-1718.
16. Shearer, G. C.; Chavan, S.; Ethiraj, J.; Vitillo, J. G.; Svelle, S.; Olsbye, U.; Lamberti, C.; Bordiga, S.; Lillerud, K. P., *Chemistry of Materials* **2014**, *26*, 4068-4071.
17. Shearer, G. C.; Forselv, S.; Chavan, S.; Bordiga, S.; Mathisen, K.; Bjørgen, M.; Svelle, S.; Lillerud, K. P., *Topics in Catalysis* **2013**, *56*, 770-782.
18. Furukawa, H.; Gandara, F.; Zhang, Y. B.; Jiang, J.; Queen, W. L.; Hudson, M. R.; Yaghi, O. M., *Journal of the American Chemical Society* **2014**, *136*, 4369-81.
19. Vandichel, M.; Hajek, J.; Vermoortele, F.; Waroquier, M.; De Vos, D. E.; Van Speybroeck, V., *CrystEngComm* **2015**, *17*, 395-406.
20. Ravon, U.; Savonnet, M.; Aguado, S.; Domine, M. E.; Janneau, E.; Farrusseng, D., *Microporous and Mesoporous Materials* **2010**, *129*, 319-329.
21. VandeVondele, J.; Krack, M.; Mohamed, F.; Parrinello, M.; Chassaing, T.; Hutter, J., *Computer Physics Communications* **2005**, *167*, 103-128.
22. Hutter, J.; Iannuzzi, M.; Schiffmann, F.; VandeVondele, J., *Wiley Interdisciplinary Reviews: Computational Molecular Science* **2014**, *4*, 15-25.
23. Ling, S.; Slater, B., *Journal of Physical Chemistry C* **2015**, *119*, 16667-16677.
24. Ling, S.; Walton, R. I.; Slater, B., *Molecular Simulation* **2015**, *41*, 1348-1356.
25. Wu, H.; Chua, Y. S.; Krungleviciute, V.; Tyagi, M.; Chen, P.; Yildirim, T.; Zhou, W., *Journal of the American Chemical Society* **2013**, *135*, 10525-32.
26. Grimme, S.; Antony, J.; Ehrlich, S.; Krieg, H., *Journal of Chemical Physics* **2010**, *132*, 154104.

27. Tschumper, G. S.; Leininger, M. L.; Hoffman, B. C.; Valeev, E. F.; Schaefer, H. F.; Quack, M., *Journal of Chemical Physics* **2002,** *116*, 690.
28. Planas, N.; Mondloch, J. E.; Tussupbayev, S.; Borycz, J.; Gagliardi, L.; Hupp, J. T.; Farha, O. K.; Cramer, C. J., *Journal of Physical Chemistry Letters* **2014,** *5*, 3716-23.
29. Yang, D.; Odoh, S. O.; Wang, T. C.; Farha, O. K.; Hupp, J. T.; Cramer, C. J.; Gagliardi, L.; Gates, B. C., *Journal of the American Chemical Society* **2015,** *137*, 7391-6.
30. Jiang, J.; Yaghi, O. M., *Chemical Reviews* **2015,** *115*, 6966-97.
31. Jasuja, H.; Peterson, G. W.; Decoste, J. B.; Browe, M. A.; Walton, K. S., *Chemical Engineering Science* **2015,** *124*, 118-124.
32. Tanaka, K.; Ito, H.; Tanaka, T., *Journal of Chemical Physics* **1987,** *87*, 1557.
33. Yang, Q.; Wiersum, A. D.; Llewellyn, P. L.; Guillerm, V.; Serre, C.; Maurin, G., *Chemical Communications* **2011,** *47*, 9603-5.
34. Ghuman, K. K.; Wood, T. E.; Hoch, L. B.; Mims, C. A.; Ozin, G. A.; Singh, C. V., *Physical Chemistry Chemical Physics* **2015,** *17*, 14623-35.
35. Stephan, D. W.; Erker, G., *Angewandte Chemie International Edition* **2010,** *49*, 46-76.